\documentclass[11pt,twoside]{article}
\usepackage{asp2010}
\usepackage{graphicx}

\resetcounters

\begin{document}

\title{Lens Inquiry: An Astronomy Lab for Non-science Majors at Hartnell Community College}
\author{Nicole~M.~Putnam$^1$, Judy~Y.~Cheng$^2$, Elizabeth~J.~McGrath$^2$, David~K.~Lai$^2$, and Pimol~Moth$^3$
\affil{$^1$Vision~Science, University~of~California, Berkeley, CA 94720}
\affil{$^2$Department~of~Astronomy~and~Astrophysics, University~of~California, Santa~Cruz, CA 95064}
\affil{$^3$Astronomy~Department, Hartnell~College, Salinas, CA 93901}}

\begin{abstract}
We describe a three hour inquiry activity involving converging lenses and telescopes as part of a semester-long astronomy lab course for non-science majors at Hartnell Community College in Salinas, CA. Students were shown several short demonstrations and given the chance to experiment with the materials, after which there was a class discussion about the phenomena they observed. Students worked in groups of 2-4 to design their own experiments to address a particular question of interest to them and then presented their findings to the class. An instructor-led presentation highlighted the students' discoveries and the lab's content goals, followed by a short worksheet-based activity that guided them in applying their new knowledge to build a simple telescope using two converging lenses. The activity was successful in emphasizing communication skills and giving students opportunities to engage in the process of science in different ways. One of the biggest challenges in designing this activity was covering all of the content given the short amount of time available. Future implementations may have more success by splitting the lab into two sessions, one focusing on converging lenses and the other focusing on telescopes.
\end{abstract}

\section{Background}\label{background}
This activity was developed as part of our participation in the Center for Adaptive Optics Professional Development Program (\citeauthor{hun08}\ \citeyear{hun08} and Hunter et al., this volume), which trains early career scientists to teach science through inquiry. Briefly, the idea behind inquiry is to teach science as real research is done: students identify their own questions, design their own experiments to answer those questions, and then share their findings with peers. We designed an inquiry-based activity for an astronomy course at Hartnell Community College to replace an existing lab about converging lenses and their application to astronomical telescopes.

\section{Venue/Audience}\label{venue}
Our activity was designed for non-science majors enrolled in a semester-long astronomy laboratory course. The lab was 2 hours and 50 minutes in length and was taught to five different sections over the course of a week with an average class size of 25 students. We assumed that the students had no content knowledge prior to the lab, and that this was predominantly their first exposure to an inquiry-based activity. 

Hartnell Community College is a Hispanic Serving Institution located in Salinas, California. Founded in 1920, Hartnell is the only public institution of higher education exclusively serving the Salinas Valley, a vast 1,000+ square mile agricultural region. The district Hartnell serves is characterized by high rates of poverty, large numbers of migrant workers, chronically high unemployment, and low educational attainment. Latinos comprise 59\% of the total enrollment at Hartnell.  Of the student population, 64\% are the first in their family to attend college.  One of the major challenges facing Hartnell is to improve the enrollment, retention, persistence, graduation, and transfer rates of its students. This is of particular concern in science, technology, engineering, and mathematics (STEM) majors where the success of underrepresented students has generally been below that of the rest of the student body. Hartnell is working toward meeting these goals through activities like the one presented here.  

\begin{figure}[!ht]
\begin{center}
\includegraphics[scale=0.7, trim=10 10 0 0]{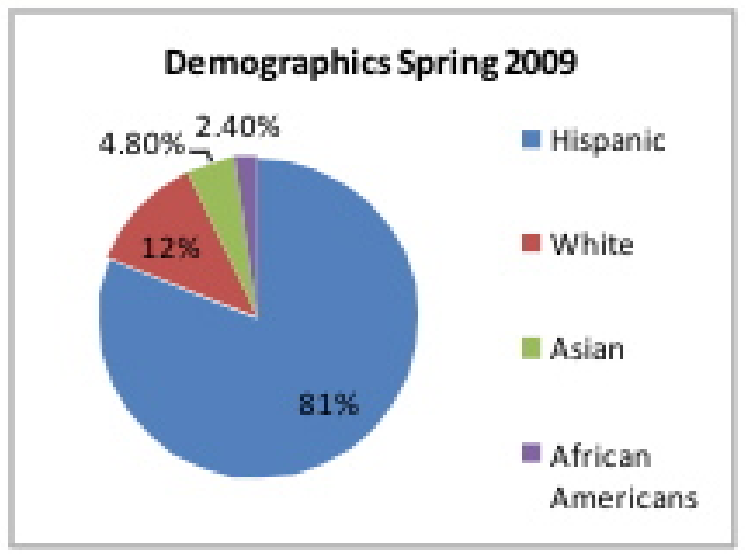}
\includegraphics[scale=0.7, trim=0 10 10 0]{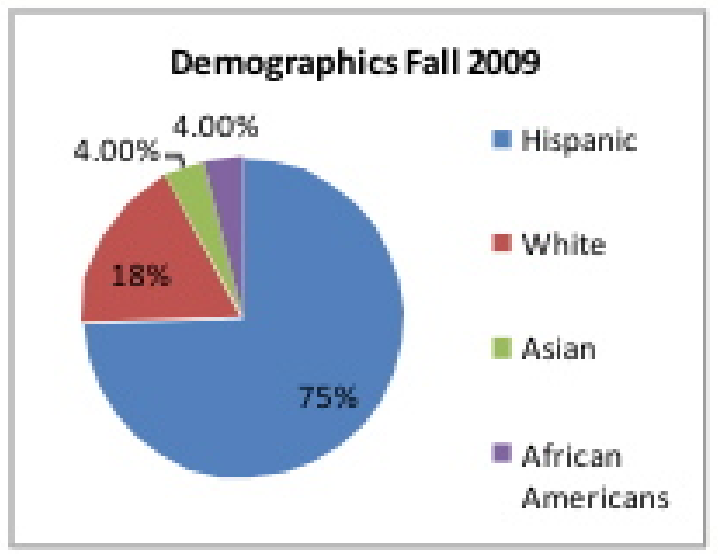}
\caption{Demographics of students enrolled in Astro-1L in 2009.}
\label{demographics}
\end{center}
\end{figure}

Figure~\ref{demographics} shows the demographics of the students who enrolled in the Introduction to Astronomy laboratory course (Astro-1L) in 2009. About 75-80\% of the students taking Astro-1L are from Hispanic backgrounds, which is significantly higher than the percentage of Hispanics enrolled at the college ($\sim$60\%). These students are historically underrepresented in the sciences and are predominantly non-science majors taking Astro-1L course to fulfill their physical sciences general education requirement. 

\section{Goals for Learners}\label{goals}
The Lens Inquiry was an introduction to inquiry for students without strong science backgrounds, so an emphasis was placed on merging scientific process goals with content at each step of the activity. We also worked to include multiple approaches to address our content goals in order to cater to multiple learning styles and engage all students in at least one aspect of the activity.  This was the first of two new inquiry-based labs that were introduced in the course, so an additional goal was to introduce the students to inquiry methods so that they could apply them again later in the semester. The second inquiry activity is described in another paper in this volume (McConnell et al.). Our goals in three different areas -- content, process, and attitudinal -- are listed below.

\subsection{Content}\label{contentgoals}
The main scientific content goals were to understand how lenses work and how their properties are used by astronomers to study distant objects. In order to limit the scope of the activity, we restricted the content to \textit{converging} lenses only. We designed the activity to emphasize the following:

\begin{itemize}
\item How lenses are used in astronomical systems to form images of distant objects
\item Different lenses have different properties intrinsic to the lens shape/material
\item Using the same lens in different ways can affect the resulting image properties
\item How a Keplerian telescope is constructed using two lenses and how it works
\end{itemize}

We divided possible converging lens investigations into five different topics: (1) lens size -- image brightness; (2) object-image distance; (3) radius of curvature; (4) image inversion; (5) lens material. 

\subsection{Inquiry Processes}
In addition to the content goals, we designed the activity to give students experience in the following science process skills (adapted from \citealp{pad90}):

\begin{itemize}
\item Planning an investigation -- Most of the other labs in this course were worksheet-based, so this was a rare opportunity for students to design their own experiments.
\item Controlling variables -- We emphasized that changing one variable at a time was a valuable tool for simplifying the problem. This was done primarily through facilitation with individual groups.
\item Communication -- We wanted to give students experience speaking in front of the class and interacting with their classmates to build their confidence and a sense of community.  We also wanted them to practice conveying their results in a way that would make sense to their peers.
\end{itemize}

\subsection{Attitudinal}
We wanted students to recognize that lenses and telescopes can be found or constructed with commonplace objects, and that specialized equipment is not necessary to appreciate the properties of lenses, although it can be beneficial to achieving consistent and accurate results in a scientific investigation. This goal was largely motivated by the results of a National Research Council report, \textquotedblleft How People Learn," \citep{don99} which discussed how students' preconceptions about how the world works affects their ability to challenge long-held beliefs and absorb new concepts. Our goal was for students to see the relevant phenomena with objects that they might come across after they left the lab so that what they learned could be continually reinforced outside of the classroom. Finally, we wanted students to have a more positive view of science and be more open to the idea of possibly pursuing science in the future.

\section{Activity Description}\label{activity}
The timeline for the Lens Inquiry is outlined in Table~\ref{description}, and detailed accounts of the activities and our goals for each section are discussed below.\footnote{\footnotesize Many more details of our lab design, including final documents such as the lab manual, telescope worksheet, and powerpoint presentations are available on our website: http://converginglenses.pbworks.com.}

\begin{table}[!ht]
\caption{Outline of Lens Inquiry Activity}
\label{description}
\smallskip
\small
\begin{center}
\begin{tabular}{ll}
\tableline
\noalign{\smallskip}
\textbf{Converging Lens Inquiry} & \textbf{2 hr 50 min}\\
\noalign{\smallskip}
\tableline
\noalign{\smallskip}
Introduction & 10 min\\
Demos/Starters & 15 min\\
Vocabulary and Question Generation & 15 min\\
Choose Question & 5 min\\
Focused Investigation/Record observations in Lab Manual & 55 min\\
Poster Making & 15 min\\
Sharing Out & 20 min\\
Converging Lenses Synthesis/Introduction to Telescopes & 10 min\\
Telescope Worksheet & 20 min\\
Telescope Synthesis & 5 min\\
\noalign{\smallskip}
\tableline
\end{tabular}
\end{center}
\end{table}

\subsection{Introduction}
Before the activity, everyone was given nametags to encourage an informal atmosphere where we could build personal relationships and a sense of community in a safe environment. The class began with the course instructor introducing the two inquiry facilitators to the students. We (the facilitators) took over from there, set up ground rules, described the process of inquiry, and discussed the timeline (Table~\ref{description}) in detail so the students knew what to expect for the duration of the activity. Finally, we introduced telescopes as motivation: (1) Telescopes are essential to astronomy research; (2) They are made of lenses; (3) We can simplify the problem by studying single converging lenses.

\subsection{Demos/Starters}
In this part of the activity, we split the class in half to see demonstrations of different aspects of the use of single lenses, which reflect the content goals we laid out when designing the activity: (1) Different lenses have different properties; (2) One lens can be used to make different images. They had approximately five minutes at each station to play with the materials themselves while recording vocabulary (any words to describe what they saw) and questions in their laboratory notebooks. Students were also encouraged to discuss their observations with each other. After finishing at one station, the groups were switched and the demonstrations were repeated.

The main goal of the demonstrations was to show the students phenomena that would generate interest. One method we used to do this was to ask students to predict what would happen at different stations, so that when something unexpected happened their interest would be piqued. For example, we illustrated image brightness by blocking light at the lens using an index card. Before introducing the index card, the students generally predicted that part of the image would disappear, and were surprised when instead, the entire image dimmed. The demonstrations were also designed to utilize everyday objects, such as magnifying glasses and drinking glasses filled with water, to show that the lens phenomena were also present in their day-to-day lives.

\subsection{Vocabulary and Question Generation, Question Selection}
In this section of the activity, students shared the vocabulary they generated during the starters with the rest of the class. One facilitator called on volunteers while the other recorded the students' responses on the board. This was useful since for many of the students in the classroom, English was not their first language.  It was also helpful in general because most students were non-science majors and were less familiar with scientific terminology. 

After the vocabulary discussion, they were asked to share questions (for example, why is the image inverted?). Discussing this as a class enabled them to think of other aspects that they may not have noticed or did not know how to describe themselves. They were instructed to add to the vocabulary and questions they recorded on their own. The primary goal was to enable a feeling of ownership over the questions that they generated, rather than requiring the students to engage in a pre-determined experiment.  The questions were recorded by a facilitator on paper strips so groups would be able to select one and take it back to their station with them. Having the questions on strips that were physically removed when chosen limited the number of groups working on the same problem, although there were still some instances of overlap. 

In each session, we aimed to cover the five main topics outlined in the content goals (\S\ref{contentgoals}). We used the vocabulary on the board as a prompt for more questions if they did not have any questions relating to a particular topic or were not speaking up.  Although we wanted to encourage all of the questions that students shared, it was necessary to restrict the class to questions that were investigable with the available materials and to re-word some questions slightly to clarify meaning.

The students were instructed to come to the front of the class where the question strips were located and to form groups based on the question they were most interested in investigating. They were required to work in groups of 2-4, and they were allowed to work with their usual lab partners if they chose to do so. 

\subsection{Focused Investigation/Record Observations in Lab Manual}
The students worked with their groups to answer their selected question. Approximately 3-4 groups were assigned to each of the facilitators for the remainder of the investigation period and the course instructor was also available as a ``floater'' to answer questions and provide additional materials as necessary. The role of the facilitators was to allow students to have ownership of their investigation (by allowing them to design and implement their own experiment) but also to confine the activity to the content goals we wanted them to focus on. This often required getting the students to focus on one particular experimental variable at a time. Logistically, we needed to allow students to work in somewhat larger groups (up to four students) to deal with the large class size. The lab manual provided a record for the instructor to go back and see what the students worked on.

\subsection{Poster Making, Sharing Out}
\begin{figure}[!ht]
\begin{center}
\includegraphics[scale=0.6]{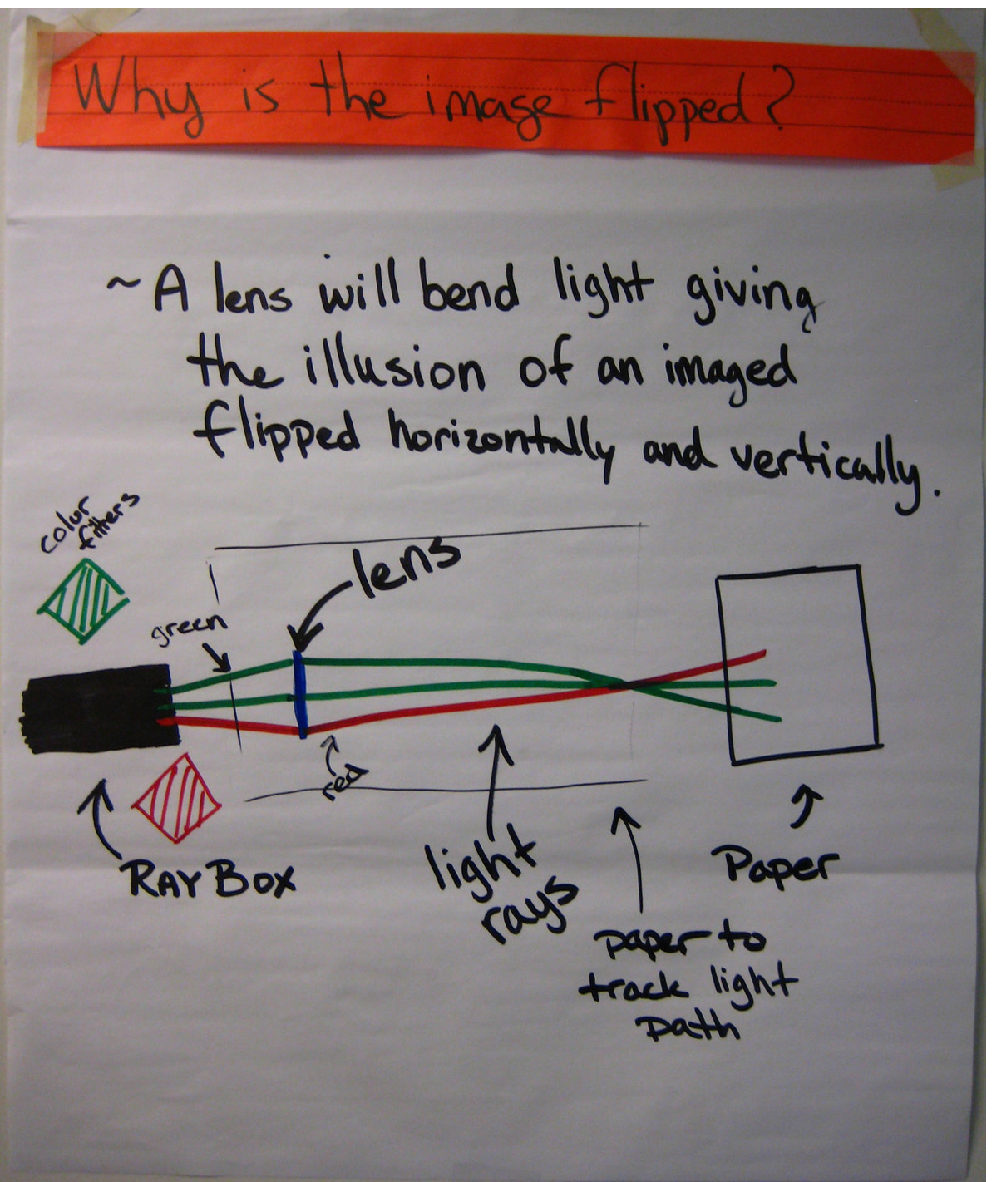}
\includegraphics[scale=0.6]{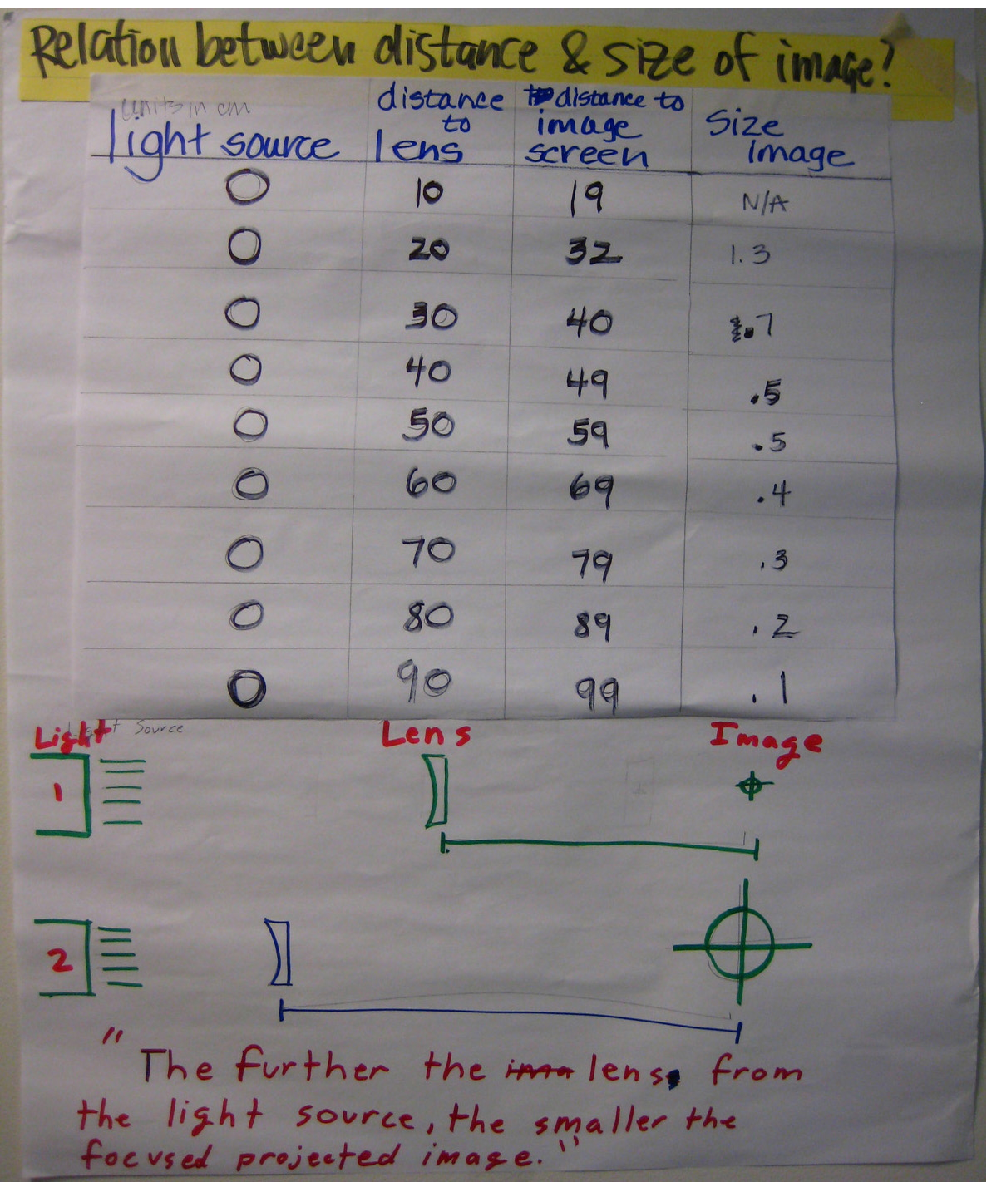}
\caption{Examples of group posters. Left: Image inversion.  Right: Image-object distance and magnification.}
\label{posters}
\end{center}
\end{figure}

Each group was required to make a poster as part of their presentation in front of the rest of the class. We suggested ways for groups to improve their posters (for example, by using diagrams), and encouraged students to think about their results using the claim-evidence-reasoning formulation \citep{mic07} which we described in the lab manual. The goal was to give students practice in communicating ideas in an understandable way. Two examples are shown in Figure~\ref{posters}.

Every group had two minutes to present their findings to the class in the style of a scientific presentation. Each group member was encouraged (but not required) to speak during the presentation and stand with their group. The purpose was to build student confidence and communication skills. This was also meant to give them an idea of how scientists conduct research; each group had an area of expertise and they shared their results with their peers who worked on other topics.

\subsection{Converging Lenses Synthesis/Introduction to Telescopes}
The main science concepts relating to converging lenses were presented to the class by a facilitator and the groups' findings were referenced in the process when possible. The goal was to make sure the class had a baseline understanding of the core concepts involving single lenses: (1) Light travels in straight lines, but it can change direction when passing through different mediums. Higher curvature/index of refraction = more bending. (2) Lenses bend light to a focal point intrinsic to that particular lens. (3) The larger the lens, the brighter the image. (4) Parallel rays get flipped beyond the focal point (images get inverted). (5) As object distance increases, image distance decreases and the image gets smaller. An introduction to the use of converging lenses in a telescope was given after the converging lens synthesis. Students were shown how to put together a simple Keplerian telescope and instructed to get the materials to construct their own. The information they learned about single converging lenses would inform their design choices when deciding which objective lens to use for their telescope.

\subsection{Telescope Worksheet, Telescope Synthesis}
The students obtained materials to construct a simple Keplerian telescope and were stepped through the process of building and using the telescope with a worksheet. Telescopes were constructed from cardboard mailing tubes and inexpensive, education-grade lenses purchased from a discount supplier.  The lenses were mounted inside the mailing tubes with weather-stripping foam tape.  We intentionally designed the telescopes to be constructed from everyday materials to further impress upon them the simplicity in the Keplerian telescope design.

Students were allowed to work in their Focused Investigation groups or choose new groups, but were still encouraged to work with at least one other person. We would have liked to give them another inquiry experience to explore how lenses are used in astronomical telescopes, but we were limited by the time constraints. We provided identical eyepiece lenses to all groups, but allowed them to choose from a variety of objective lenses. The goal was in part to give them more ownership of the activity, but mostly to show them that different lenses result in different telescope properties. For example, a large lens results in a brighter image, but the telescope will be larger and harder to mount. The main idea was to get them thinking about the different considerations that go into designing state-of-the-art telescopes. As a part of this section, each group consulted with another group to compare and contrast their telescope design and use. The discussions allowed students to interact with their classmates, some of whom they had not spoken to before, and it also gave them an additional telescope to examine. 

We concluded with a brief synthesis to summarize the activity and explain how this would relate to their upcoming visit to Fremont Peak Observatory. In the synthesis we compared two large refracting telescopes to mirror the part of their activity where they compared their telescope with another group.

\section{Assessment}\label{assessment}

\subsection{Formative Assessment}
One facilitator was assigned to each group of 2-4 students, with an average of 2-4 groups per facilitator. We attempted to limit the number of facilitators in order to make the activity sustainable in the future, for a course with only the instructor and one teaching assistant. However, as it was our first time implementing the activity, we had two facilitators working with the course instructor at each class. In its current form, this is the minimum number we would recommend in order to give students enough guidance to reach the content goals in the time allotted. 

The role of the facilitators was to answer questions as they arose, but also to assess whether the groups were on track. One way to test the students' understanding was to see if they could explain a new piece of information. For example, many groups that had questions involving the shape of the lens believed that the thickness of the lens was important, but could not explain what they saw when they used a thick lens with no curvature. They were considered successful if they could make an accurate prediction or explain why their inaccurate predictions were wrong.

\subsection{Summative Assessment}
The main summative assessment of the activity was the written record from each student. This included the lab manual where students recorded their questions and observations, the telescope worksheet, and the group posters (see Figure~\ref{posters}).

\begin{figure}[!htb]
\begin{center}
\includegraphics[scale=0.71,trim=25 200 25 25, angle=90]{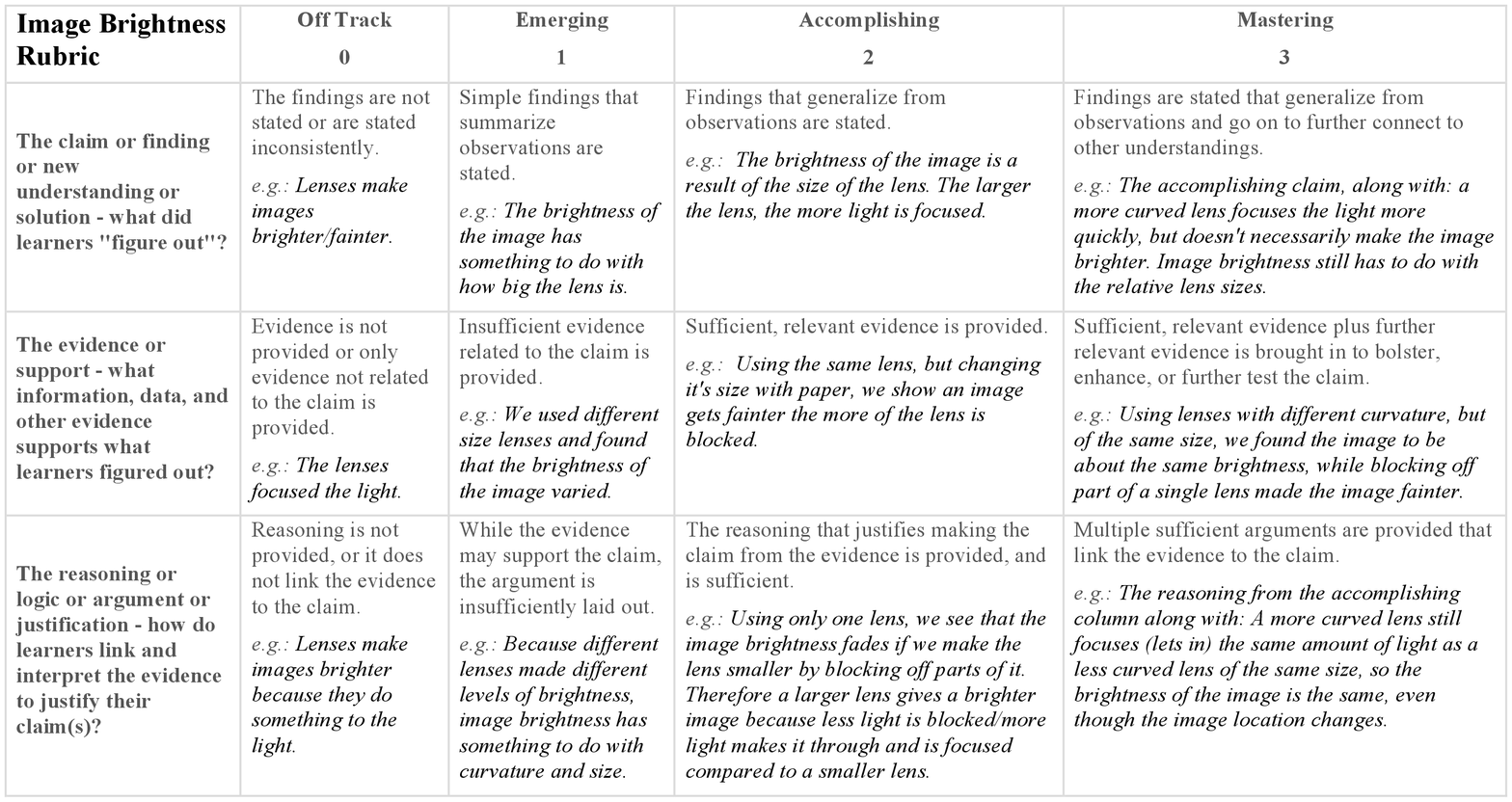}
\caption{Example rubric used for assessment.}
\label{rubric}
\end{center}
\end{figure}

In addition to this written summative assessment, we developed rubrics prior to the activity to determine what we considered basic information that the students should be able to explain; an example for image brightness is shown in Figure~\ref{rubric}. We conferred about what kind of understanding would qualify as \textquotedblleft Emerging," \textquotedblleft Accomplishing," and \textquotedblleft Mastering." These rubrics were useful in the design of our activity and during facilitation to remind us of the content goals for each topic of investigation. 

We scored groups during their presentations using the rubrics, but this was difficult for a number of reasons. Because the presentations were very short, it was difficult to fill out the rubric in the time allotted and record notes to include in the synthesis. We also concluded that overall our content goals were perhaps too high given the knowledge level of the students and the short time that was available for the investigation.  

Lastly, the students were asked a question on their final exam that tested their knowledge of converging lenses, \textquotedblleft When you block the left half of the object, what image should you see on the other side of a converging lens?" This was one of the harder, more conceptual questions on the exam. Even so, 58\% of the students got the correct answer, which was close to the average of the lab final.

\section{Social and/or Cultural Aspects of the Design in Practice}\label{socialcultural}
In designing our activity, the non-science background of the students was a driving consideration, especially since this would likely be their first exposure to inquiry. We wanted the students to feel engaged in the activity, but  we worried that the content was not particularly exciting to the general public. Though we did not expect that we could try to change their minds about pursuing science careers, we did not want them to feel alienated by specialized lab equipment. Therefore, we made a concerted effort to use materials that related to daily life -- normal light bulbs, magnifying glasses, drinking glasses filled with water and sugar water -- in both the starter demonstrations and the investigations. We could then make the link to more sophisticated laboratory equipment if they were attempting to make a quantitative measurement that would be aided by its use, but this was not always necessary.

One of the main issues we encountered was emphasizing the content while keeping them engaged. It seemed that the groups that were more interactive did not necessarily get all of the content goals and the opposite was also happening. Our design intentionally offered opportunities for the learners to engage and learn in different ways. In general, we felt that the structure of the activity, in which we put students in many different situations where they could observe, experiment, and interact with each other, allowed most students to feel engaged during at least one part of the activity, and this was reflected in many of the students' positive post-activity feedback.

Given the time constraints and the variety of stations and activities, however, we had to move them quickly from section to section, which prevented some students from getting the science content from different parts of the activity. Furthermore, perhaps because of their non-science backgrounds, many of the groups settled early on the \textquotedblleft answers" to their questions, and ignored any evidence that their explanations might not be correct. For example, some groups who investigated the effect of lens shape determined that the thickness of a lens changed the kind of image that was observed (when in fact it was the curvature), but ignored the thick flat lens that did not form an image.

\section{Considerations for the Future}\label{future}
We believe the Lens Inquiry was a good introduction to inquiry for community college students in an astronomy course. It builds off a basic scientific concept and offers opportunities to engage a variety of student learning styles. Time constraints and our desire to incorporate an application to astronomy with the addition of a telescope worksheet led to aspects of the activity feeling rushed.  It also created a demand for more facilitators than may be available in many community college settings. This is an important consideration for instructors planning adaptations of the activity for their own classrooms. We now highlight a few aspects of the activity that could be revised and a few aspects that we feel are particularly strong and should not be overlooked.

The following is a list of revisions that may improve future implementations:
\begin{enumerate}
\item \textbf{Time Constraints:} The biggest challenge by far was including all of the content into a three hour period. Many students complained that the activity was rushed and there was too much to do. We would suggest dividing this activity into two inquiries -- (1) Converging Lenses and (2) Telescopes. One of the original goals was for the instructor to be able to do this with one other person. It does not seem feasible considering the class sizes. Dividing the activity into two sessions and limiting some of the available materials and possible investigations may help.
\item \textbf{Poster Making:} Many groups needed more time to accomplish this task than anticipated and a more focused approach/guide may have been useful. Many groups wanted to explain everything that they tried in the investigation period rather than the key content concepts they discovered. We may want to emphasize what makes a good poster, the claim-evidence-reasoning formulation, and how to distill the most important facts. This could be done with a focused class discussion or poster/presentation templates that the students could refer to. 
\item \textbf{Materials:} We may want to further limit the amount of materials and questions that students can investigate in the future. The effect of changing the lens material was the most challenging topic to investigate, perhaps because the only available materials (glasses filled with sugar water) were not carefully crafted lenses.
\item \textbf{Ray boxes:} We often found that the groups working with image inversion, who were provided with ray boxes that allowed them to trace light rays through the lens, got a better understanding of what caused the phenomenon they were studying. Initially, we worried that the ray boxes would give too much information away, but seeing how it benefited some of the groups made us reconsider. Ray boxes would need to be incorporated into the starters at the beginning.
\item \textbf{Rubrics:} In retrospect, some of the expectations, especially for the \textquotedblleft Mastering" category, were quite high.
\end{enumerate}

Despite these difficulties, our activity was very successful in getting students to build their communication skills. The class discussions were active and generally produced a range of questions for groups to investigate. Most of the students seemed relaxed during their presentations. In addition, students were able to interact with each other more informally through discussions about their observations during the starters and talking to their classmates about how their telescopes compared. One student commented that the activity helped them to get to know their classmates. Finally, the students seemed comfortable with the facilitators, even though they had never met before, and were open about asking questions during the focused investigations.

In addition, since this was the first inquiry activity, the students' experience here made the second inquiry (McConnell et al., this volume) run more smoothly, as some of the difficulties with this activity were taken into account. In particular, students were allowed more time to work on their investigations and poster presentations. The students also were familiar with inquiry and knew more about what to expect, so they were more prepared and likely managed their time better during the second activity.

Overall, the Lens Inquiry was an improvement over the existing lab because it got students thinking for themselves and gave them more of a hands-on experience compared to their typical guided labs. The activity would need to be revised to make it feasible for just an instructor and one facilitator, by splitting it into two sessions and by further limiting the materials and investigations. 

\acknowledgements The authors wish to thank Lynne Raschke and Lisa Hunter for help with the design and implementation of the activity.  This material is based upon work supported by: the National Science Foundation (NSF) Science and Technology Center program through the Center for Adaptive Optics, managed by the University of California at Santa Cruz (UCSC) under cooperative agreement AST\#9876783; NSF DUE\#0816754; and the UCSC Institute for Scientist \& Engineer Educators.  N.\ M.\ P.\ acknowledges funding from the AOF William C. Ezell Fellowship. D.\ K.\ L. acknowledges the support of the National Science Foundation through the NSF AAPF program under award AST\#0802292.

\bibliographystyle{asp2010}
\bibliography{putnam}

\end{document}